\documentclass[showpacs,twocolumn,prb]{revtex4-2}
\usepackage{amsfonts}
\usepackage{graphicx}
\usepackage{amsmath}
\usepackage{amssymb}

\begin{document}
\title{Hexagonal warping on optical conductivity of surface states in
Topological Insulator $Bi_{2}Te_{3}$}
\author{Zhou Li$^1$}
\email{lizhou@univmail.cis.mcmaster.ca}
\author{J. P. Carbotte$^{1,2}$}

\affiliation{$^1$ Department of Physics, McMaster University,
Hamilton, Ontario,
Canada,L8S 4M1 \\
$^2$ Canadian Institute for Advanced Research, Toronto, Ontario,
Canada M5G 1Z8}

\begin{abstract}
ARPES studies of the protected surface states in the Topological Insulator $%
Bi_{2}Te_{3}$ have revealed the existence of an important hexagonal warping
term in its electronic band structure. This term distorts the shape of the
Dirac cone from a circle at low energies to a snowflake shape at higher
energies. We show that this implies important modifications of the interband
optical transitions which no longer provide a constant universal background
as seen in graphene. Rather the conductivity shows a quasilinear increase
with a slightly concave upward bending as energy is increased. Its slope
increases with increasing magnitude of the hexagonal distortion as does the
magnitude of the jump at the interband onset. The energy dependence of the
density of states is also modified and deviates downward from linear with
increasing energy.
\end{abstract}

\date{\today }
\maketitle

\section{Introduction}

Topological Insulators are insulating in the bulk and have symmetry
protected helical Dirac fermions on their surface with an odd number of
Dirac points in the surface state Brillouin zone.\cite{Zhang, Fu1, Balents,
Moore, Hsieh1, Chen1, Hsieh2, Hsieh3, Kane} The surface charge carriers are
massless and relativistic with linear in momentum energy dispersion curves.
Spin sensitive angular resolved photo emission spectroscopy (ARPES) also
shows that their spins are locked to their momentum. The position of the
chemical potential relative to the Dirac point and the gapped bulk bands is
not as easily tuned as it is in graphene \cite{Geim} but can be controlled
by doping with $Sn$ in $(Bi_{1-\delta}Sn_{\delta})_{2}Te_{3}$ or with $Ca$
in $Bi_{2-\delta}Ca_{\delta}Te_{3}$ with further dosing with $NO_{2}$
molecules. The constant energy contours in $Bi_{2}Te_{3}$ as measured by
ARPES are not circles as they are in graphene but have an hexagonal
distortion which gives them snowflake shape. This geometry was modeled by Fu
\cite{Fu2} with an unconventional hexagonal warping term in the bare band
Hamiltonian of $Bi_{2}Te_{3}$ with parameters fit to the measured Fermi
surface.

Optical spectroscopy has been a very powerful method to obtain valuable
information on the charge dynamics of the Dirac fermions in graphene. \cite%
{Carbotte1, Carbotte2,Li} An experimental review was given by Orlita and
Potemski.\cite{Orlita} Usually it is the zero momentum q limit of the
optical conductivity as a function of photon energy which is measured but
very recently finite q's have also been measured using near field
techniques. \cite{Fei1,Fei2,Chen2,Nicol} Optics has also been used to study
topological insulators. \cite{Hancock,Basov,Homes} In this paper we study
how hexagonal warping in $Bi_{2}Te_{3}$ manifest in the optical
conductivity, which we find is profoundly modified for the parameters
determined in the work of Fu.\cite{Fu2}

\section{Formalism}

The Hamiltonian used by Fu \cite{Fu2} to describe the surface states band
structure near the $\Gamma $ point in the surface Brillouin zone is
\begin{equation}
H_{0}=v_{k}(k_{x}\sigma _{y}-k_{y}\sigma _{x})+\frac{\lambda }{2}%
(k_{+}^{3}+k_{-}^{3})\sigma _{z}+E_{0}(k)  \label{Hamiltonian}
\end{equation}
where $E_{0}(k)=\hbar ^{2}k^{2}/(2m^{\ast })$ is a quadratic term which
gives the Dirac fermionic dispersion curves an hour glass shape and provides
particle-hole asymmetry. The Dirac fermion velocity to second order is $%
v_{k}=v_{F}(1+\alpha k^{2})$ with $v_{F}$ the usual Fermi velocity measured
to be $2.55eV\cdot \mathring{A}$ and $\alpha $ is a constant which is fit
along with $m^{\ast }$ to the measured band structure in the reference [%
\onlinecite{Fu2}]. The hexagonal warping parameter $\lambda =250eV\cdot
\mathring{A}^{3}$. The $\sigma _{x}$, $\sigma _{y}$, $\sigma _{z}$ are the
Pauli matrices here referring to spin, while in graphene these would relate
to pseudospin instead. Finally $k_{\pm }=k_{x}\pm ik_{y}$ with $k_{x}$, $%
k_{y} $ momentum along $x$ and $y$ axis respectively. The energy spectrum
associated with the Hamiltonian [Eq.~(\ref{Hamiltonian})] is
\begin{equation}
E_{\pm }(k)=E_{0}(k)\pm \sqrt{v_{k}^{2}k^{2}+\lambda ^{2}k^{6}\cos
^{2}(3\theta )}  \label{disp}
\end{equation}
where $\theta $ is the polar angle defining the direction of $k$ in the two
dimensional surface state Brillouin zone. The energy dispersion curves in
Eq.~(\ref{disp}) reduce to the well known linear law $\pm v_{k}k$ of
graphene when $E_{0}$ is set zero along with $\lambda =0$, i.e. ignoring
hexagonal warping. Since our primary interest here is getting a first
understanding of how the warping term in Eq.~(\ref{disp}) manifests itself
in the dynamical conductivity of the surface helical Dirac fermions we will
for simplicity from here on drop the $E_{0}(k)$ term which as we have said
provides particle-hole asymmetry.
\begin{figure}[tp]
\begin{center}
\includegraphics[height=3.0in,width=3.0in,angle=-90]{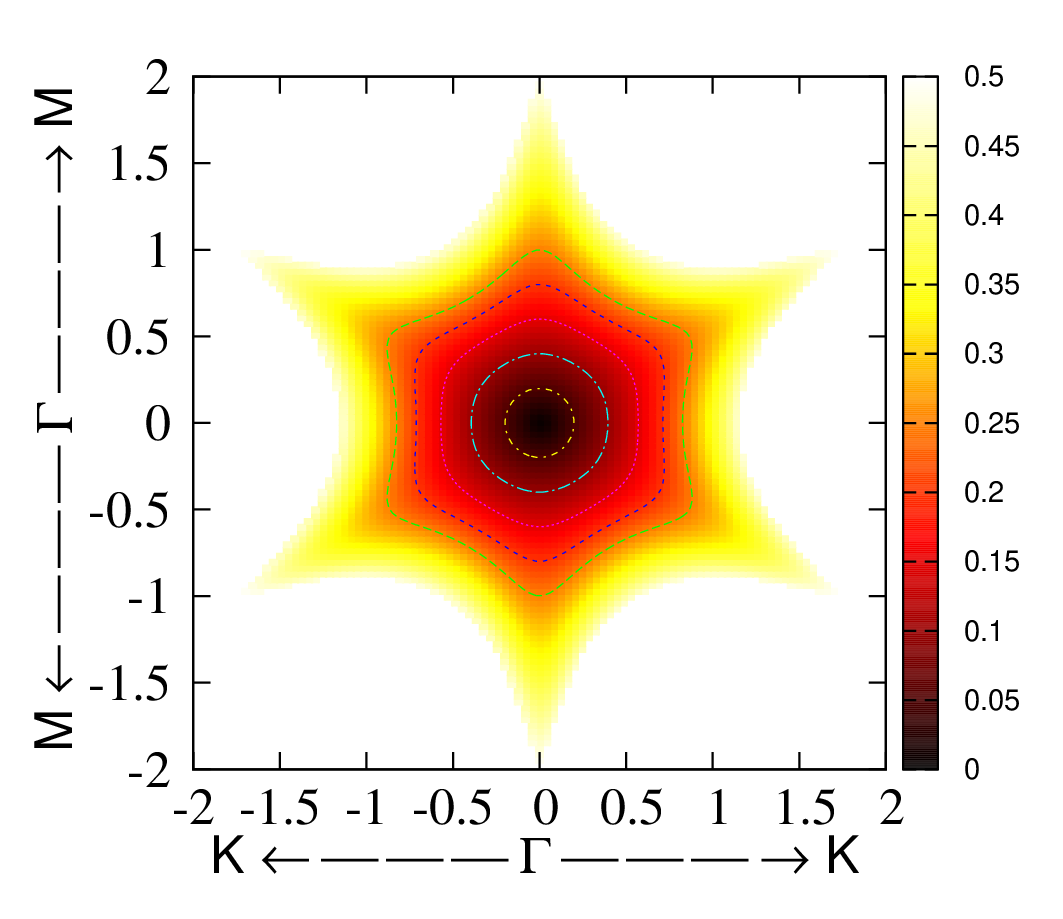} %
\includegraphics[height=2.0in,width=2.0in]{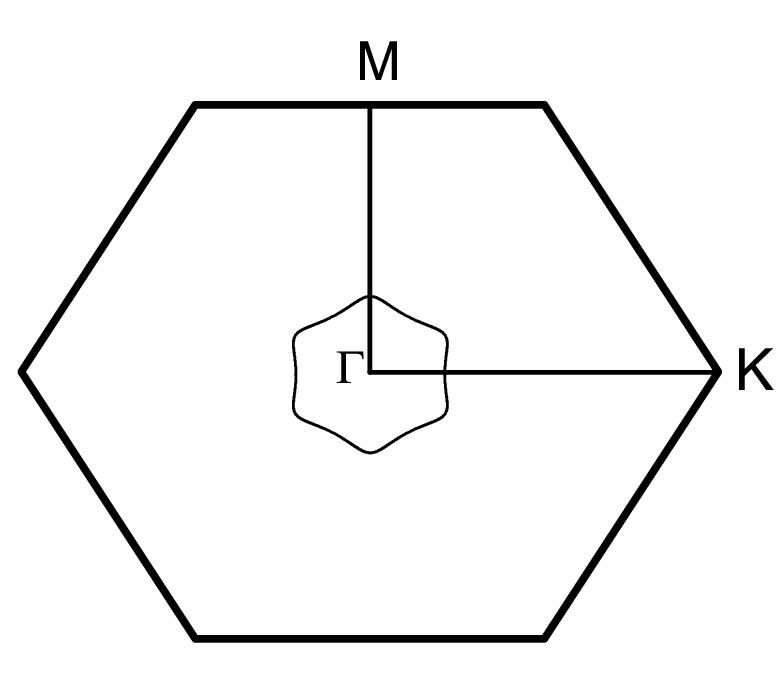}
\end{center}
\caption{(Color online) Constant energy contours for the dispersion curves
used to describe the bare bands in $Bi_{2}Te_{3}$ with chemical potential $%
\protect\mu$ changed by doping with $Sn$ in $(Bi_{1-\protect\delta}Sn_{%
\protect\delta})_{2}Te_{3}$ where $\protect\delta =.67\%$ corresponds to a
chemical potential $\protect\mu =250meV$. The $kx$ and $ky$ axes are in the
units of $0.1\mathring{A}^{-1}$. Also shown is the surface state Brillouin
zone identifying $\Gamma,K$ and $M$ points.}
\label{fig1}
\end{figure}

In FIG.~\ref{fig1} we show a color plot for the constant energy contour
associated with the dispersion curves [Eq.~(\ref{disp})] as the energy is
increased above that of the Dirac point the contour changes shape and shows
greater hexagonal distortion displaying a snowflake shape. The largest flake
shown corresponds to a chemical potential of 250meV which is achieved when $%
\delta =0.67\%$ in $(Bi_{1-\delta }Sn_{\delta })_{2}Te_{3}$. In graphene we
would have circles for all energies and in strained graphene we would have
elliptical contours instead of circle.

The Kubo formula for the $xx$ component of the dynamic conductivity $\sigma
_{xx}(\omega )$ as a function of photon energy $\omega $ is given in terms
of the matrix Matsubara Green's function $\hat{G}(k,\omega _{n})$ with $%
\omega _{n}$ the Fermionic Matsubara imaginary frequency as%
\begin{eqnarray}
&&\sigma _{xx}(\omega )=-\frac{e^{2}}{i\omega }\frac{1}{4\pi ^{2}}%
\int_{0}^{k_{cut}}kdkd\theta  \notag \\
&&T\sum_{l}Tr\langle v_{x}\hat{G}(\mathbf{k,}\omega _{l})v_{x}\hat{G}(%
\mathbf{k,}\omega _{n}+\omega _{l})\rangle _{i\omega _{n}\rightarrow \omega
+i\delta }
\end{eqnarray}%
with $e$ the charge on the electron, $k$ the absolute value of the momentum
with direction $\theta $ and $k_{cut}$ a cut off. Here $T$ is the
temperature with $\omega _{n}=(2n+1)\pi T$ and $\omega _{l}=2l\pi T$ the
Fermion and Boson Matsubara frequencies,\cite{Mahan} $n$ and $l$ are
integers and $Tr$ is a trace. To get the conductivity which is a real
frequency quantity, we needed to make an analytic continuation from
imaginary $i\omega _{n}$ to real $\omega $ and $\delta $ is infinitesimal.
As written we have neglected vertex correction and so the factors $v_{x}$
are simply the velocity components given by
\begin{eqnarray}
v_{x} &=&v_{k}\sigma _{y}+3\lambda k^{2}\cos (2\theta )\sigma _{z}
\label{vx} \\
v_{y} &=&-v_{k}\sigma _{x}-3\lambda k^{2}\sin (2\theta )\sigma _{z}
\label{vy}
\end{eqnarray}%
obtained directly from the Hamiltonian [Eq.~(\ref{Hamiltonian})]. We have
set all $\hbar $ factors equal to one.

The matrix Green's function for the non-interacting bare band is given by
\begin{eqnarray}
&&\hat{G}_{0}(\mathbf{k},i\omega _{n})  \notag \\
&=&\frac{1}{i\omega _{n}+\mu -v_{k}(k_{x}\sigma _{y}-k_{y}\sigma _{x})-\frac{%
\lambda }{2}(k_{+}^{3}+k_{-}^{3})\sigma _{z}}  \notag \\
&=&\frac{i\omega _{n}+\mu +v_{k}(k_{x}\sigma _{y}-k_{y}\sigma _{x})+\lambda
k^{3}\cos (3\theta )\sigma _{z}}{(i\omega _{n}+\mu
)^{2}-v_{k}^{2}k^{2}-\lambda ^{2}k^{6}\cos ^{2}(3\theta )}
\end{eqnarray}
with%
\begin{equation}
(k_{+}^{3}+k_{-}^{3})^{2}=(2k_{x}^{3}-6k_{x}k_{y}^{2})^{2}=4k^{6}\cos
^{2}(3\theta )
\end{equation}%
It is convenient to rewrite the $\hat{G}_{0}(\mathbf{k},i\omega _{n})$ in
terms of $\hat{G}_{0}(\mathbf{k},s,i\omega _{n})$ defined as%
\begin{equation}
G_{0}(\mathbf{k},s,i\omega _{n})=\frac{1}{i\omega _{n}+\mu -s\sqrt{%
v_{k}^{2}k^{2}+\lambda ^{2}k^{6}\cos ^{2}(3\theta )}}
\end{equation}

where $s=\pm 1$ and $\mathbf{F}_{k}$ defined as
\begin{equation}
\mathbf{F}_{k}=\frac{(-v_{k}k\sin \theta ,v_{k}k\cos \theta ,\lambda
k^{3}\cos (3\theta ))}{\sqrt{v_{k}^{2}k^{2}+\lambda ^{2}k^{6}\cos
^{2}(3\theta )}}
\end{equation}%
This gives%
\begin{equation}
\hat{G}_{0}(\mathbf{k},i\omega _{n})=\frac{1}{2}\sum_{s=\pm }(1+s\mathbf{F}%
_{k}\cdot \mathbf{\sigma })G_{0}(\mathbf{k},s,i\omega _{n})
\end{equation}

\section{Simplification of expression for $\protect\sigma _{xx}(\protect%
\omega )$}

We will be interested here only with the interband terms in which case the
required trace gives%
\begin{eqnarray}
&&Tr\langle v_{x}\hat{G}(\mathbf{k,}\omega _{l})v_{x}\hat{G}(\mathbf{k,}%
\omega _{l}+\omega _{n})\rangle  \notag \\
&=&\frac{H(\theta )}{[v_{k}^{2}k^{2}+\lambda ^{2}k^{6}\cos ^{2}(3\theta )]}
\notag \\
&&(\frac{1}{i\omega _{n}+\mu -E_{-}}\frac{1}{i\omega _{n}+i\omega _{l}+\mu
-E_{+}}  \notag \\
&&+\frac{1}{i\omega _{n}+\mu -E_{+}}\frac{1}{i\omega _{n}+i\omega _{l}+\mu
-E_{-}})
\end{eqnarray}%
where%
\begin{eqnarray}
H(\theta ) &=&\lambda ^{2}v_{k}^{2}k^{6}[\cos ^{2}(3\theta )-2\times3\cos \theta
\cos (2\theta )\cos (3\theta )  \notag \\
&&+9\cos ^{2}(2\theta )]+v_{k}^{4}k^{2}\sin ^{2}\theta
\end{eqnarray}
But we know that
\begin{eqnarray}
&&T\sum_{l}[\frac{1}{i\omega _{l}+\mu -E_{-}}\frac{1}{i\omega _{l}+i\omega
_{n}+\mu -E_{+}}  \notag \\
&&+\frac{1}{i\omega _{l}+\mu -E_{+}}\frac{1}{i\omega _{l}+i\omega _{n}+\mu
-E_{-}}]  \notag \\
&=&[\frac{f(E_{-})-f(E_{+})}{i\omega _{n}-E_{+}+E_{-}}+\frac{%
f(E_{+})-f(E_{-})}{i\omega _{n}-E_{-}+E_{+}}]
\end{eqnarray}%
and hence we obtain
\begin{eqnarray}
&&\sigma _{xx}(\omega )=-\frac{e^{2}}{i\omega }\frac{1}{4\pi ^{2}}%
\int_{0}^{k_{cut}}kdkd\theta \frac{H(\theta )}{[v_{k}^{2}k^{2}+\lambda
^{2}k^{6}\cos ^{2}(3\theta )]}  \notag \\
&&[\frac{f(E_{-})-f(E_{+})}{i\omega _{n}-E_{+}+E_{-}}+\frac{f(E_{+})-f(E_{-})%
}{i\omega _{n}-E_{-}+E_{+}}]_{i\omega _{n}\rightarrow \omega +i\delta }
\end{eqnarray}%
Here $f(x)$ is the Fermi-Dirac distribution function given by $f(x)=1/[\exp
(x/T-\mu /T)+1]$, where we have ignored the Boltzman constant $k_{B}$ but
will include it in the calculation. We have verified that $\sigma
_{yy}(\omega )=\sigma _{xx}(\omega )$, after an analytic continuation from
imaginary to real Matsubara frequencies we obtain the final expression%
\begin{eqnarray}
&&\sigma _{xx}(\omega )=-\frac{e^{2}}{i\omega }\frac{1}{4\pi ^{2}}%
\int_{0}^{k_{cut}}kdkd\theta \frac{H(\theta )}{W(k,\theta)}\times  \notag \\
&&[\frac{f(E_{-})-f(E_{+})}{\omega -2\sqrt{W(k,\theta)}+i\delta }+\frac{%
f(E_{+})-f(E_{-})}{\omega +2\sqrt{W(k,\theta)}+i\delta }]  \label{sxx}
\end{eqnarray}
where $W(k,\theta)=v_{k}^{2}k^{2}+\lambda ^{2}k^{6}\cos ^{2}(3\theta )$.

\section{Analytic form for the real part of conductivity}

The real part of the dynamic conductivity which is the absorptive part can
be simplified further using the usual rule $\frac{1}{\omega +i\delta }=\frac{%
P}{\omega }-i\pi \delta (\omega )$. From Eq.~(\ref{sxx}) we get%
\begin{eqnarray}
&&Re\sigma _{xx}(\omega )=-\frac{e^{2}}{\omega }\frac{1}{4\pi ^{2}}%
\int_{0}^{k_{cut}}kdkd\theta H(\theta )\times  \notag \\
&&\frac{\lbrack f(E_{-})-f(E_{+})]}{[W(k,\theta )]}(-\pi )\delta (\omega -2%
\sqrt{W(k,\theta )})
\end{eqnarray}%
which can be rewritten as%
\begin{eqnarray}
&&Re\sigma _{xx}(\omega )=\frac{e^{2}}{2\omega }\frac{1}{4\pi }%
\int_{0}^{k_{cut}}d(k^{2})d\theta \frac{H(\theta )}{\omega ^{2}/4}\times  \notag \\
&&[f(-\omega /2)-f(\omega /2)]\delta (\omega -2\sqrt{W(k,\theta )})
\end{eqnarray}%
and we get%
\begin{eqnarray}
&&Re\sigma _{xx}(\omega )=\frac{e^{2}[f(-\omega /2)-f(\omega /2)]}{24\pi
\omega ^{2}}  \notag \\
&&\times \int_{0}^{k_{cut}}\frac{d(k^{2})H(\theta _{k,\omega })}{|\lambda
^{2}k^{6}\cos (3\theta _{k,\omega })\sin (3\theta _{k,\omega })|}
\end{eqnarray}%
where the thermal factor $f(\omega )$ have been pulled out of the integral
over momentum. The Dirac delta function has been used in the process. We
wrote%
\begin{equation}
\delta (\omega -2\sqrt{W(k,\theta )})=\frac{\omega \delta (\theta
-\theta _{k,\omega })}{|12\lambda ^{2}k^{6}\cos (3\theta _{k,\omega
})\sin (3\theta _{k,\omega })|}
\end{equation}

where%
\begin{eqnarray}
&&\theta _{k,\omega }=\pm \frac{1}{3}\arccos [\pm \sqrt{\frac{\omega
^{2}/4-v_{k}^{2}k^{2}}{\lambda ^{2}k^{6}}}],  \notag \\
&&\pm \frac{1}{3}\{\arccos [\pm \sqrt{\frac{\omega ^{2}/4-v_{k}^{2}k^{2}}{%
\lambda ^{2}k^{6}}}]+\pi \},  \notag \\
&&\pm \frac{1}{3}\{\arccos [\pm \sqrt{\frac{\omega ^{2}/4-v_{k}^{2}k^{2}}{%
\lambda ^{2}k^{6}}}]+2\pi \}.  \label{angle}
\end{eqnarray}%
This is our final expression for the absorptive part of the conductivity.
Numerical results based on this expression are given in the next section.
Before doing so however we present a similar expression for the density of
states $D(\omega )$ as a function of energy which could be measured in
scanning tunneling microscopy (STM). By its definition%
\begin{equation}
D(\omega )=-\frac{1}{\pi }\sum_{\mathbf{k}}ImTr\hat{G}(\mathbf{k},i\omega
_{n}->\omega +i\delta )
\end{equation}%
which can be reduced to%
\begin{eqnarray}
&&D(\omega )=-\frac{1}{\pi }\frac{1}{4\pi ^{2}}\int_{0}^{k_{cut}}kdkd\theta
\notag \\
&&Im[\frac{1}{\omega -\sqrt{W(k,\theta )}+i\delta }+\frac{1}{\omega +\sqrt{%
W(k,\theta )}+i\delta }]
\end{eqnarray}%
Taking the imaginary part gives%
\begin{eqnarray}
D(\omega ) &=&\frac{1}{4\pi ^{2}}\int_{0}^{k_{cut}}kdkd\theta \delta (\omega
-\sqrt{W(k,\theta )})\Theta (\omega ) \notag\\
+(\omega  &\rightarrow &-\omega )
\end{eqnarray}%
which works out to
\begin{eqnarray}
D(\omega ) &=&\frac{\omega }{12\pi ^{2}\lambda ^{2}}\int_{0}^{k_{cut}}\frac{%
kdk}{|k^{6}\cos (3\theta _{k,\omega }^{\prime })\sin (3\theta _{k,\omega
}^{\prime })|}\Theta (\omega )  \notag \\
+(\omega  &\rightarrow &-\omega )
\end{eqnarray}%
where $\Theta (\omega )$ is the Heaviside step function and we have used
\begin{equation}
\delta (\omega -\sqrt{W(k,\theta )})=\frac{\omega \delta (\theta
-\theta _{k,\omega }^{\prime })}{|3\lambda ^{2}k^{6}\cos (3\theta
_{k,\omega }^{\prime })\sin (3\theta _{k,\omega }^{\prime })|}
\end{equation}%
where $\theta _{k,\omega }^{\prime }$ can be obtained from Eq.~(\ref{angle})
by replacing $\omega $ with $2\omega $ so we have $\theta _{k,\omega
}^{\prime }=\theta _{k,2\omega }$.

\section{Numerical Results}

\begin{figure}[tp]
\begin{center}
\includegraphics[height=3.0in,width=3.0in]{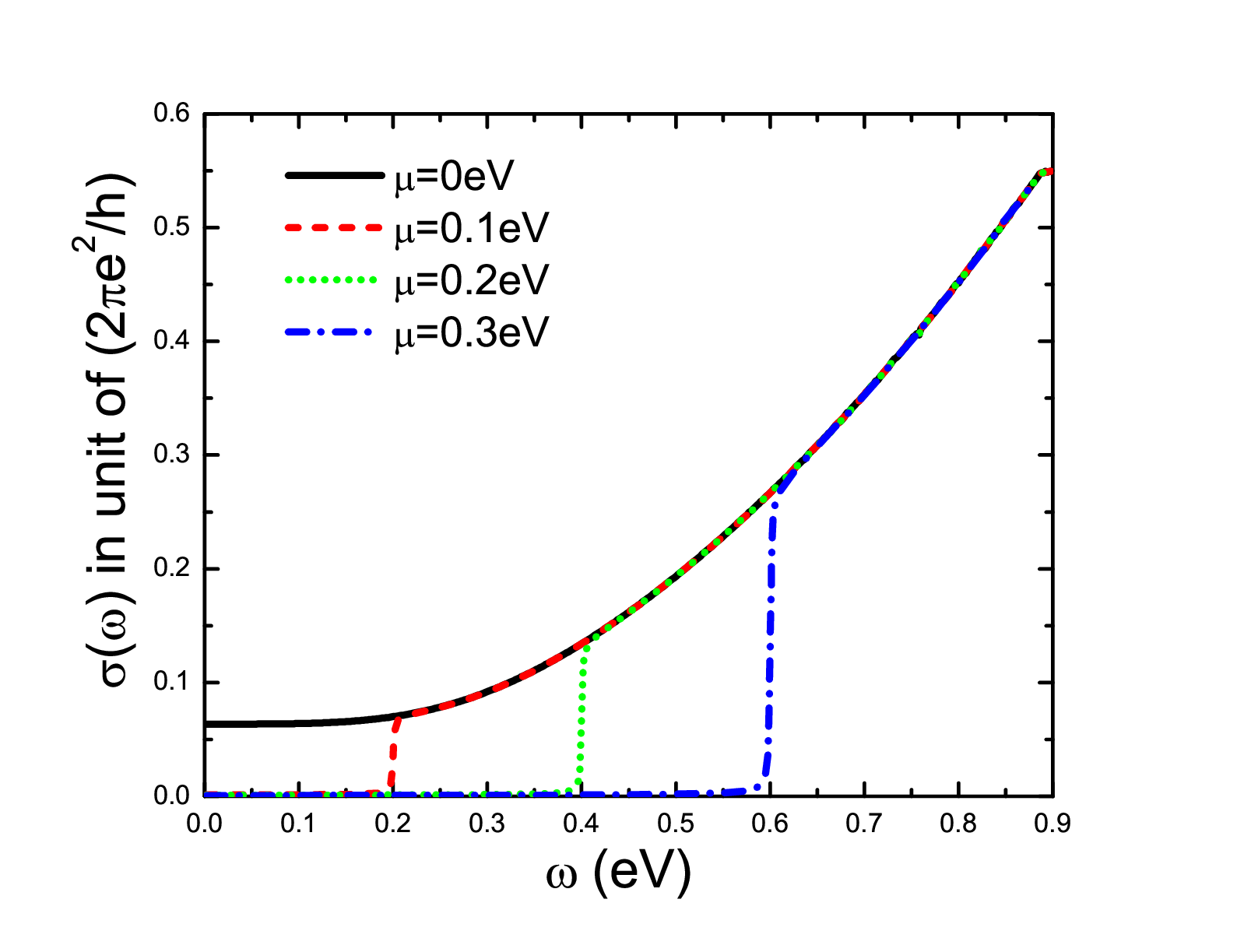}
\end{center}
\caption{(Color online) The real part of the optical conductivity $\protect%
\sigma _{xx}(\protect\omega )$ as a function of photon energy $\protect%
\omega $ in meV (in units of $2\protect\pi e^{2}/h$) for a case which
corresponds approximately to $\protect\delta=0.67\%$ $Sn$ doping with
chemical potential $\protect\mu=0.25eV$. We show 4 values of $\protect\mu$.
In all cases finite $\protect\mu$ transfers optical spectral weight from the
interband transitions to the intraband. These, not shown here, provide a
Drude like contribution at $\protect\omega$ near zero.}
\label{fig2}
\end{figure}

\begin{figure}[tp]
\begin{center}
\includegraphics[height=4.0in,width=4.0in]{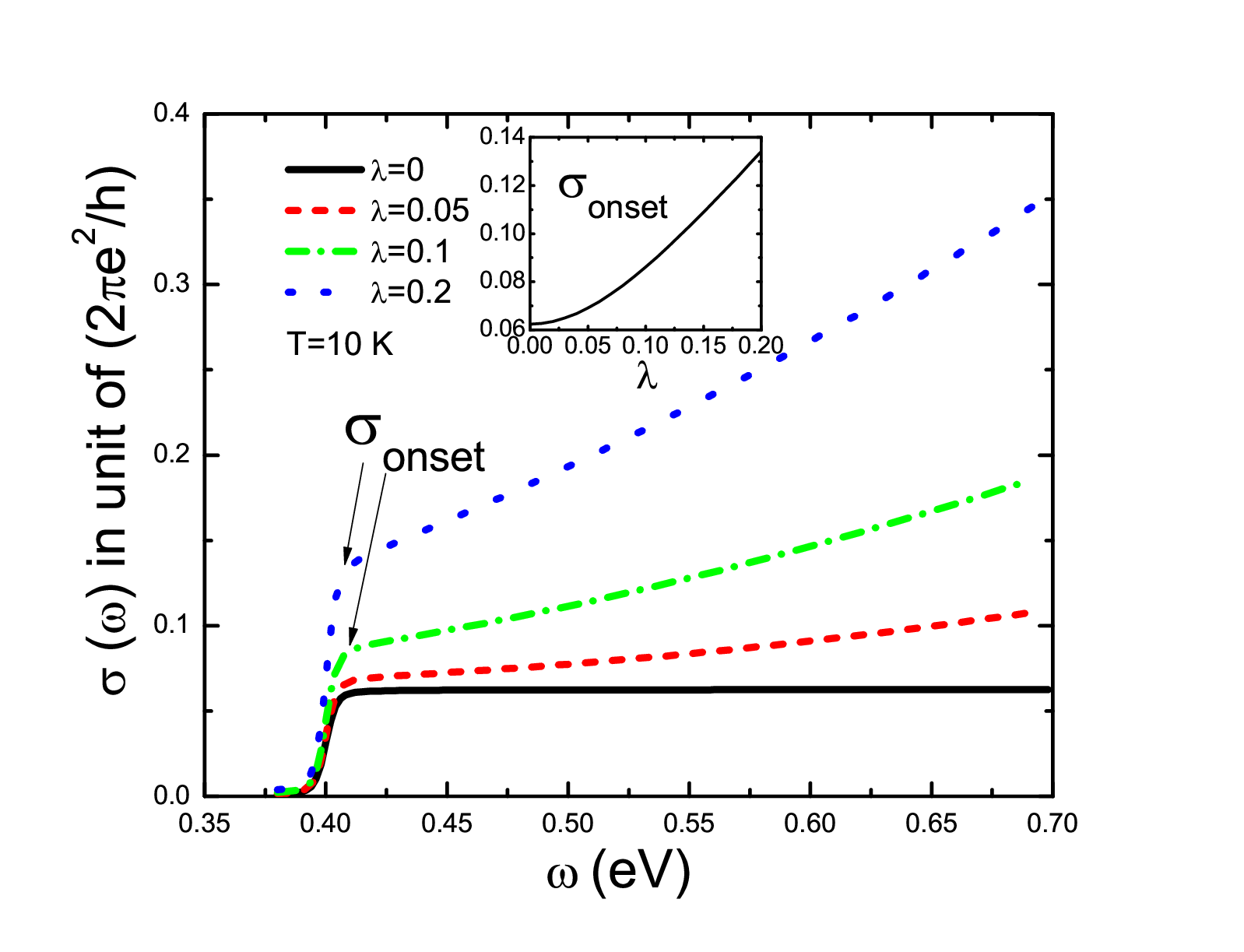}
\end{center}
\caption{(Color online) The real part of the optical conductivity $\protect%
\sigma _{xx}(\protect\omega )$ in units of $2\protect\pi e^{2}/h$ as a
function of photon energy for 4 values of the magnitude of the hexagonal
warping term ($\protect\lambda$) in the Hamiltonian (1). The solid black
curve is for comparison. In this case $\protect\lambda=0$ and the model
reduces to the contribution to the optics of a single spin and single valley
Dirac cone of graphene. In the inset we show the increase in the value of
the jump at the interband absorption edge with increasing $\protect\lambda$.}
\label{fig3}
\end{figure}

\begin{figure}[tp]
\begin{center}
\includegraphics[height=3.0in,width=4.0in]{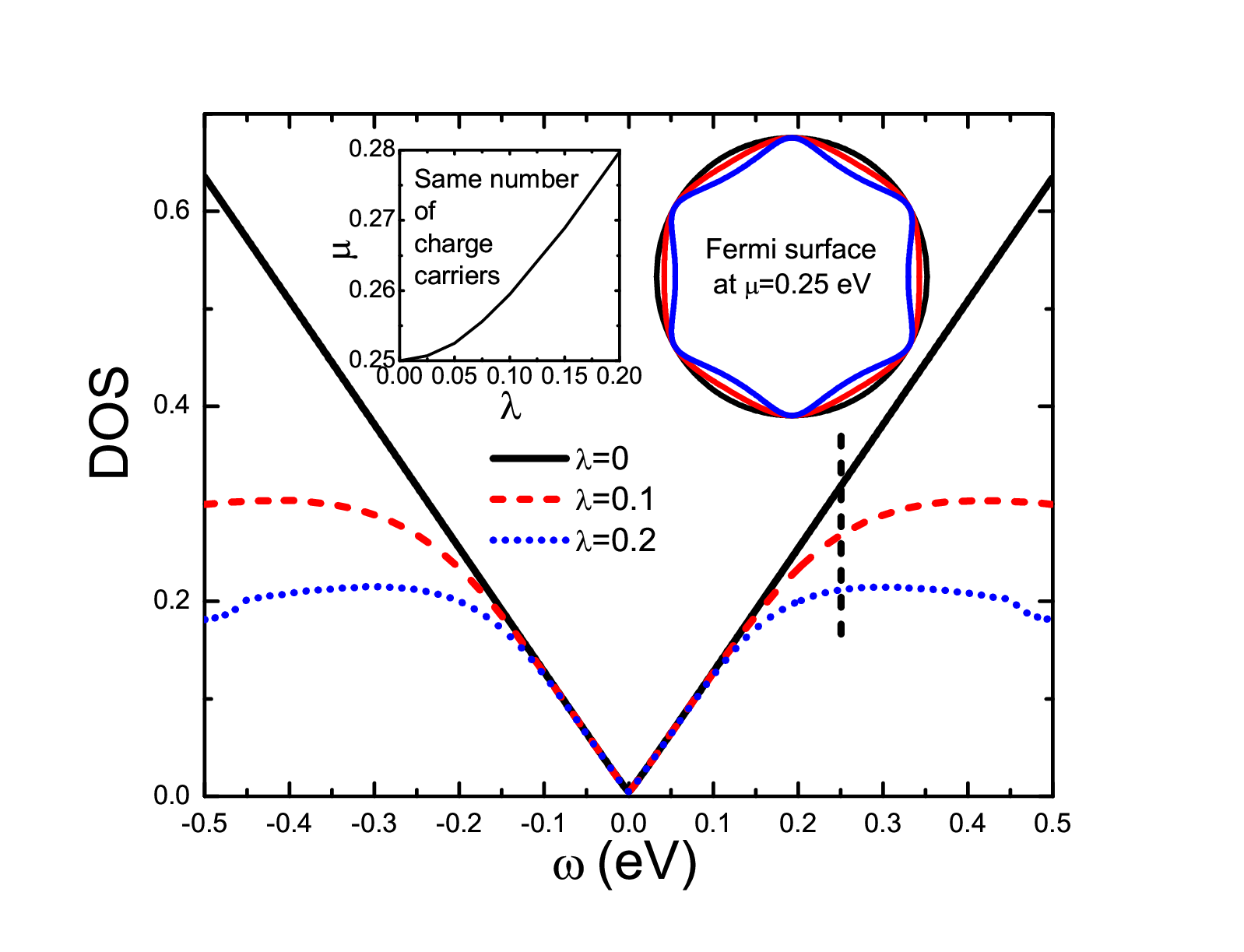}
\end{center}
\caption{(Color online) The density of state $D(\protect\omega)$ as a
function of $\protect\omega$. We show these for values of the hexagonal
distortion term $\protect\lambda=0$ (solid black line), $\protect\lambda=0.1$
(dashed red line) and $\protect\lambda=0.2$ (dotted blue line). The solid
black straight line corresponds to graphene without spin and valley
degeneracy. The Fermi velocity $v_{F}$ was taken to be that appropriate to $%
Bi_{2}Te_{3}$. The vertical dashed black line indicates a chemical potential
$\protect\mu=0.25eV$. The left inset gives the value of $\protect\mu$ as a
function of $\protect\lambda$ for fixed number of charge carriers which is
around $1.1*10^{16}/m^{2}$. The right inset shows the Fermi surface at fixed
value of $\protect\mu$ for $\protect\lambda=0$ (black), $\protect\lambda=0.1$
(red), $\protect\lambda=0.2$ (blue). With no hexagonal distortion it is
circular and distorts to a snowflake as $\protect\lambda$ increases.}
\label{fig4}
\end{figure}

In FIG.~\ref{fig2} we show our results for the real part of the optical
conductivity $Re\sigma _{xx}(\omega )$, in units of $2\pi e^{2}/h$, as a
function of photon energy $\omega $ for a $Bi_{2}Te_{3}$ doped with $Sn$ at
level $\delta =.67\%$ (see references (\onlinecite{Chen1}) and (%
\onlinecite{Fu2})) which corresponds to a chemical potential $\mu =250meV$
and all other parameters determined in the fit by Fu.\cite{Fu2} We show four
values of $\mu $. In all cases the threshold for the start of the interband
transitions is sharp and occurs at $\omega =2\mu $, as it would in graphene.
The missing optical spectral weight in the interband transition is
accompanied with an increase in the intraband (Drude) optical spectral
weight. This is not shown in our picture. Because we have not included any
scattering processes in our work, the Drude manifests as a Dirac delta
function at $\omega =0$ and does not overlap with the interband contribution
which we emphasize here. At small values of $\omega $, the value of $%
Re\sigma (\omega )$ is rather flat and takes on precisely the value expected
for graphene without the degeneracy factor $g=4$, which counts spin and
valley degrees of freedom. For a topological insulator there is only one
Dirac cone and spin is no longer degenerate. We also note that the
background value is independent of material parameters such as the Fermi
velocity. But this is no longer the case for a topological insulator. As $%
\omega $ is increased whatever the value of $\mu $ the conductivity
increases rather rapidly above its universal background value and shows
concave upward behavior. This is traced to the changes in fermi velocity of
Eq.~(\ref{vx}) and Eq.~(\ref{vy}) due to the warping term proportional to $%
\lambda $ and to the change in quasiparticle band structure. In FIG.~\ref%
{fig3} we show how $Re\sigma (\omega )$ v.s. $\omega $ is changed as $%
\lambda $ is changed. Here and also in FIG.~\ref{fig4} the $\lambda $ has
been multiplied by the cube of the typical Fermi momentum, which is $0.1
\mathring{A}^{-1}$. So $\lambda=0.2$ here would correspond to $%
\lambda=200eV\cdot \mathring{A}^{3} $. For reference the case
$\lambda =0$ given as the solid black curve, corresponds precisely
to graphene except for value of the degeneracy factor $g$. In this
case the universal background, well known in the graphene
literature, is recovered. As $\mu$ is increased the optical spectral
weight lost in the background is transferred to the intraband
contribution not shown above. Increasing $\lambda $ changes the band
structure and no optical sum rule applies. As $\lambda $ is
increased at fixed value of chemical potential $\mu $ the magnitude
of the interband onset ($\sigma _{onset}$) (which remains at $2\mu
$) increases as shown in the inset. In addition its increase at
$\omega>2\mu $ becomes even more rapid. While in the black dashed
curve it is reasonably linear, the dotted curve for $\lambda =0.2$
has acquired a significant upward curvature. These deviations from
the universal background of graphene is the signature in optics of
the hexagonal warping term in our Hamiltonian
[Eq.~(\ref{Hamiltonian})]. We have checked and found that these
curves do not scale onto each other. The increase in $Re\sigma
(\omega )$ with $\omega $ above the universal background value is to
be contrasted with the behavior of the quasiparticle density of
states $D(\omega )$ given by Eq. (24). As in graphene, the case
$\lambda =0$ gives $D(\omega )$ proportional to $|\omega | $. As
$\lambda $ is increased however $D(\omega )$ starts to deviate from
linearity and, as we see in FIG.~\ref{fig4}, is progressively
reduced below the solid black curve. This is easily understood with
the help of the right inset where we plot the constant energy
contours for $\omega =\mu =0.25eV$ for the three value of $\lambda $
considered. For $\lambda =0$ we get the black circle of graphene
theory. As $\lambda $ increases this contour distorts into a
snowflake pattern (blue curve) which is however completely contained
inside the black circle. Of course, to keep the number of charge
carriers the same we need to increase the chemical potential with
increasing value of the warping parameter $\lambda $ as we show in
the left inset. What is plotted is the value of $\lambda $ at fixed
value of the number of charge carriers, which is around $1.1\ast
10^{16}/m^{2}$.

\section{Summary and Conclusion}

Helical Dirac fermions exist at the surface of a topological insulator (TI).
These charge carriers have some similarity and also differences with the
well known chiral Dirac fermions in graphene. An important difference is a
degeneracy factor $g$ of four which comes from the valley and spin degrees
of freedom of graphene not applicable in TI. Another important difference,
well investigated in the case of $Bi_{2}Te_{3}$ doped with $Sn$, is the
hexagonal distortion seen in ARPES. Here we have studied how such a term
changes optical properties. For realistic values of the warping parameter we
found large changes in the interband transitions.

A third difference is that, graphene involves pseudospin related to the
sublattice degeneracy in its two atoms per unit cell crystal structure
rather than real spin. Furthermore in graphene the bands are spin
degenerate, while in a topological insulator momentum and spin are locked
with x-y component of real spin oriented perpendicular to its 2-D momentum
k, with clockwise and anticlockwise orientation in conduction and valence
band respectively.

The universal flat background observed in graphene\cite{Carbotte2} remains
at small photon energies although modified by a factor of 4 because the
valley spin degeneracy no longer applies. As $\omega $ increases large
modifications in the effect of the interband transitions on the conductivity
are noted, and these encode the information on the hexagonal warping of the
Dirac cone cross-section leading to a snowflake pattern. Instead of being
flat $Re\sigma (\omega )$ increases in a quasilinear fashion with a concave
upward bent. The magnitude of this linear increase becomes larger with the
magnitude of the hexagonal warping term as does the value of the jump in the
conductivity at the threshold of twice the chemical potential ($\omega =2\mu
$). At the same time, we find that the density of state remains linear only
at small $\omega $ and starts to fall below this linear behavior at the
energy where the conductivity also starts to show its deviation from a
constant background value. While the conductivity curves are bent upward due
to fermi velocity features, the density of state bends downward a prediction
that could be verified in combined optics and scanning tunneling
spectroscopy(STS) experiment.

\begin{acknowledgments}
This work was supported by the Natural Sciences and Engineering
Research Council of Canada (NSERC) and the Canadian Institute for
Advanced Research (CIFAR).
\end{acknowledgments}

\end{document}